# Investigating the effects of nanoparticles and surfactants on heat transfer performance of water in nucleate pool boiling experiment

M. Abdelrashied, *H. Yousery*, M. Roshdy, M. Metwally, M. Ali, H. Elsawy, *R. Saqr*
Benha University – Shoubra faculty of engineering

*Abstract--*The main objective of the present study is to investigate the effect of adding nanoparticles at different concentrations and sizes to boiling distilled water on the nucleate pool boiling heat transfer performance. The major studied parameters are the effects of wall heat flux, type of nanoparticles, particle size, and its concentration in aqueous solutions. The tested nanoparticles are silica (SiO2) and alumina (Al2O3) nanoparticles at 0.001, 0.002, 0.003, 0.005, and 0.01 vol% concentrations, and particle sizes of 10, 20, and 50nm. A pool boiling test rig is designed and constructed to conduct heat transfer experiments with Arudino Control System and eight Type-K thermocouples for temperature measurements. The experimental investigations showed that adding definite amounts of nanoparticles to the water at a particular concentration and size improves its physical and thermal properties and consequently improves the heat transfer coefficient in percentages reaching 70% for SiO2. For Al2O3, it showed noticeable lower values compared to Al2O3; however, equal to distilled water. Furthermore, the surfactants showed slightly better HTC values than distilled water which diminishes at high heat flux values.

## Nomenclature

| | |
|---|---|
| CHF | Critical Heat Flux |
| HTC | Heat Transfer Coefficient |
| Ni-Cd | Nickel–cadmium |
| CTAB | Cetyl Trimethyl Ammonium Bromide |
| SLES | Sodium Laureth Sulfate |
| SDS | Sodium dodecyl sulfate |

## I. Introduction

Boiling heat transfer is considered one of the most important cooling methods used in domestic and commercial refrigeration and air conditioning systems. In addition to refrigeration and air conditioning applications, conventional and enhanced liquids have a wide range of industrial applications, including power generation, chemical and petrochemical production, metallurgical quenching process, electronics cooling, and desalination seawater, in nuclear power plants and either of heating or electricity generation (Ciloglu D, 2005). Boiling at the surface of a body immersed in an extensive pool of stagnant liquid is called "pool boiling". Boiling is probably the most familiar form of heat transfer, yet it remains the least understood form.

A *nanofluid* is a fluid that contains nanoparticles, which are nanometer-sized particles. These fluids are colloidal nanoparticle suspensions in a base fluid that has been designed. Metals and oxides are commonly utilized as nanoparticles in nanofluids (Kamatchi R, 2005). Water and oil are all common base fluids.

Using nanofluids in nucleate pool boiling is an important research topic as it offers improvements in HTC and CHF. Such enhancements are vital for making boiling systems more energy-efficient to reduce energy consumption and system size and volume. Furthermore, enhancing critical heat flux (CHF) is vital to make boiling systems compact and safe to endure high heat fluxes.

## II. Literature review

(Thome, 1990), (Webb, 1994), and (Bergles, 1997) documented several active and passive techniques used for pool boiling enhancement. Amongst these, the use of additives, which lower the surface tension of liquids. Small amounts of certain surfactant additives change the boiling phenomena (Wasekar and Manglik, 2000) drastically. Such boiling phenomena have received continuous interest for a long time. In the experiments of (Selim, 2006) Ni-Cd alloys have been electrodeposited from a solution containing cationic CTAB, anionic SLS, and non-ionic Triton X-100 to improve the surface quality of the specimens. (Manglik, 2009) investigated the pool boiling behavior experimentally over a smooth cylindrical heating surface made from copper. He represented the enhancement in heat transfer by the relative heat transfer coefficient.

A nanofluid, first reported by (Choi, 1995), is a new class of fluids engineered by dispersing nanometer-sized solid particles (1–100 nm) in the base fluids. Nanofluid boiling began to draw research interest in 2003 and has become an important research area of improving boiling fluids' thermal properties. (Kamatchi and Venkatachalapathy, 2015) conducted a review on the pool boiling CHF enhancement with nanofluids. The review was organized according to parameters, including surfactants, the material, size and concentration of nanoparticles, thermal transport properties, system pressure, and nanoparticle deposit layer. The review only cited one related paper dated in 2014 and none after that.

(Kshirsagar and Shrivastava, 2015) reviewed the influence of nanoparticles on thermal conductivity and pool boiling HT and CHF.

(Wu and Zhao, 2013) reviewed nanofluid studies in the thermophysical properties, convective HT performance, boiling HT performance, and CHF enhancement, and all articles cited were dated in or before 2012. The (Ahn and Kim, 2011) review included articles in the CHF of pool boiling and flow boiling of nanofluids, and all cited articles were dated in or before 2011. (Celen et al. 2014) reviewed the literature related to Nano refrigerants, including flow and pool boiling. A Nano refrigerant is a nanofluid with a refrigerant as the base fluid. (Bahiraei and Hangi, 2015) reviewed the investigations of magnetic nanofluids (MNFs), and only several articles investigating the pool boiling, and flow boiling were cited. An MNF is a nanofluid comprised of a non-magnetic base fluid and magnetic nanoparticles, in which the fluid flow, particle movement, and heat transfer process can be controlled by applying magnetic fields.

### III. Experiment

A test rig is designed, fabricated, and constructed in which nucleate pool boiling of aqueous surfactant solutions is admitted to the electrically heated horizontal copper tube to achieve the requirements mentioned above. The test rig is installed at the air-conditioning laboratory, Faculty of Engineering at Shoubra, Benha University.

#### A. Formulation and characterization of nanofluids

we investigated silica (SiO2) and alumina (Al2O3) nanoparticles at 0.001, 0.002, 0.003, 0.005, and 0.01 vol% concentrations, and particle sizes of 10, 20, and 50nm. These particles were developed and supplied to our lab by "Housing & Building National Research Center" in Cairo, Egypt.

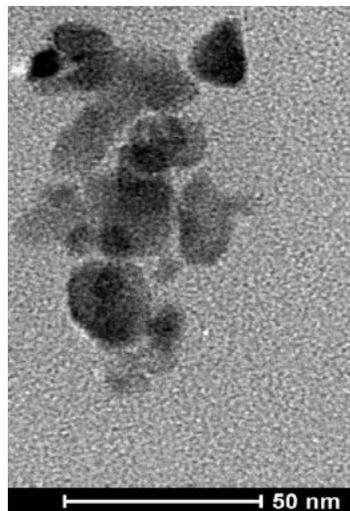

Fig. 1. Exemplary SiO2 nanoparticles at 50nm size.

Two main techniques are used for producing Nanofluids, one- and two-step methods. The first represents the direct formation of the nanoparticles in a base fluid. The second represents the preparation of nanoparticles separately mixed into the base fluid. In either case, pH control, surfactant, and ultrasonic vibration can be applied to achieve a stable suspension. Compared to fluids containing milli- or micro-sized particles, Nanofluids exhibit superior heat transfer performance and better stabilization. The main reasons are higher surface area and higher thermal conductive capability than the base fluid.

The experimental procedure starts by breaking down the agglomeration. The time varies according to the concentration and usually takes between five to ten minutes. Afterward, the entrained gases are removed from the layers of the fluid using a piece of special equipment for degassing liquid. It takes up to five minutes for optimal results. Finally, the nanoparticles are dispersed in the fluid and suspended through vibration using an ultrasonic vibrator under temperature 30°C. It takes 45 minutes/liter. These steps occur sequentially and not in parallel. Therefore, preparing one liter of nanofluid takes around 90 minutes.

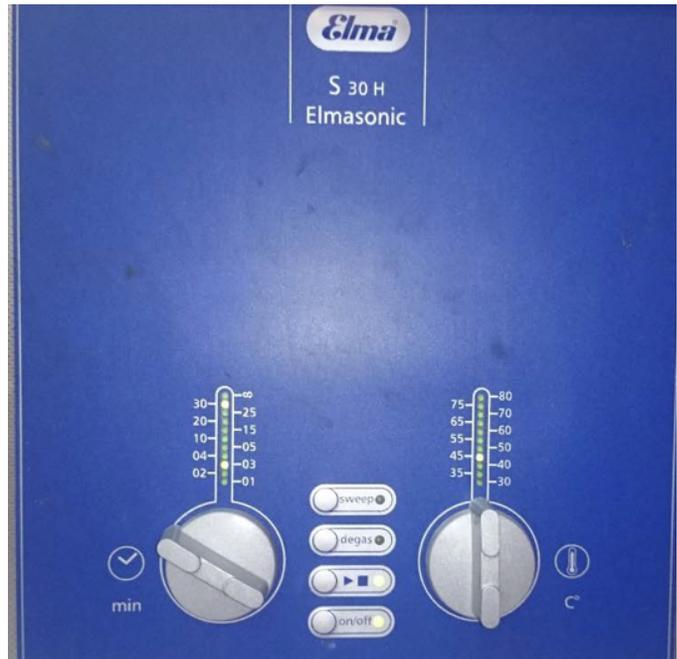

Fig. 2. An ultrasonic vibrator device was used to sweep, degassing, and suspend the particles in the fluid of the experimental test facility.

#### B. Boiling experimental apparatus

A line diagram of the experimental setup for the present study is shown in Fig. 3, while Fig. 4 shows a photograph of the experimental test facility. These two figures show that the apparatus consists of the following main components: boiling and condensation vessel, test tube (heated surface), condenser cooling water circuit, and instrumentation.

The test fluid (either distilled water or aqueous surfactant solution with a specific concentration) evaporates on the outer surface of an electrically heated horizontal copper tube within the boiling and condensation vessel. The vapor rises to a shell and coil condenser where the circulating cooling water removes the latent heat of vaporization. The condensate returns to the liquid pool by gravity. Experiments are conducted under the condition of a saturated nucleate pool boiling at an operating pressure of 1 bar.

The steady-state condition within the test loop is achieved by matching the heat removed at the condenser coil to be equal to the power input of the electrically heated surface of the test tube. It can be accomplished by adjusting the difference between the saturation temperature of the test liquid and the inlet cooling water temperature by controlling the cooling water flow rate. In the next section, complete details of the main components of the experimental apparatus will be presented.

Fig. 3 shows a schematic diagram of the boiling and condensation vessel. It consists of two stainless steel (316 L) hollow cylinders of different diameters connected linearly by a semi-conical shape reducer of 80 mm height. The upper cylinder is used for condensation, and the lower cylinder is used as the evaporation section. The evaporation section is provided with two glass windows (90 mm diameter each) perpendicular to each other at a level of 70 mm from the bottom. These two glass windows are used to observe bubble formation at the test tube's visual surface and photographic recordings of the boiling process. The vessel bottom is provided with two ports; the first one supports the auxiliary heater (2 kW heating capacity) and accordingly provides additional heating/stirring during the degassing process before test runs. The second port is matched with a gate valve used to charge the test fluid and drainage vessel. Besides, the vessel's upper flange is provided with one port that branched into two branches. The first branch is matched with a calibrated pressure gauge (-1: 5 bar) to measure the operating pressure. The second branch is used to charge the system with compressed air (4 bar) for 24 hours to detect any leak within the system or evacuate it before charging with the test fluid.

tube represents a tight-fit cavity for the immersible electric cartridge heater. The cartridge heater (12 mm diameter, 100 mm long, 220V, 1kW) with insulated wires is press-fitted with conductive grease into a copper tube casing of 22 mm, outer diameter. Two stainless steel caps encapsulate this assembly at both ends to minimize axial heat loss. The saturation temperature of the selected test fluid is measured by eight calibrated copper-constantan thermocouples located at different positions inside the pool and in the vapor space. The use of the cooling water circuit aims to achieve two purposes. The first is to release the latent heat of condensation of the ascending vapor. The second is to maintain the constant saturation temperature of the aqueous solution by accurately adjusting the heat removed by the cooling water passing through the condenser coil to the heat produced at the test tube boiling surface.

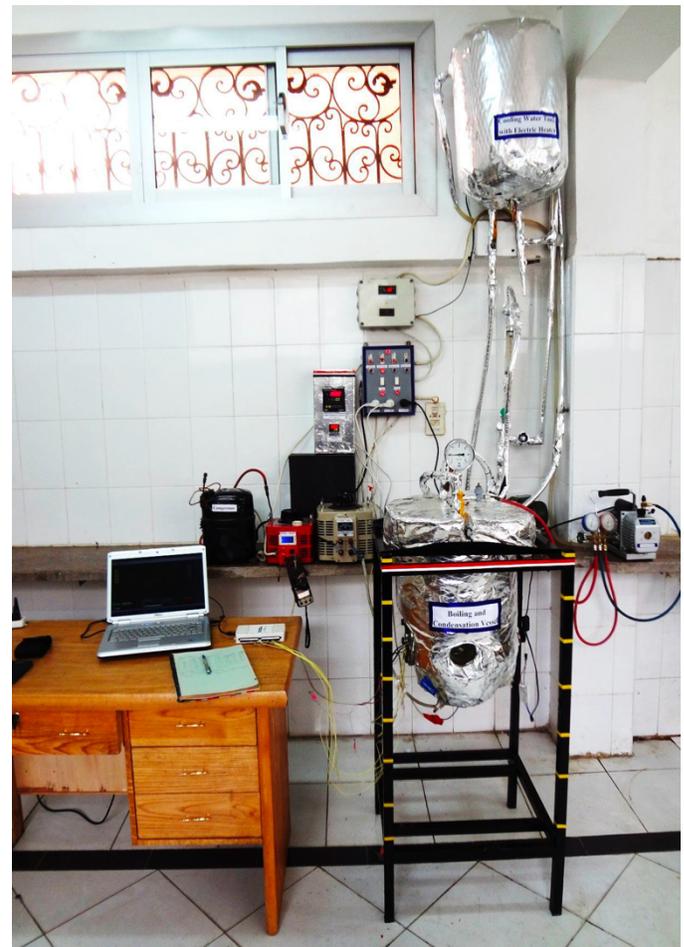

Fig. 4. Photograph of the experimental test facility.

## C. Arduino Control System

Suitable measuring instruments are used for monitoring voltage drop, electric current, temperature, and volume flow rate to accomplish the present experimental study. The following items give the specifications of the measuring instruments. A digital multimeter (Model: A, M3800) is used to measure the difference in the input voltage drop across the heater terminals with accuracy + 0.1 V. Two A. C. voltage regulators are used in the current test rig. The first is used to

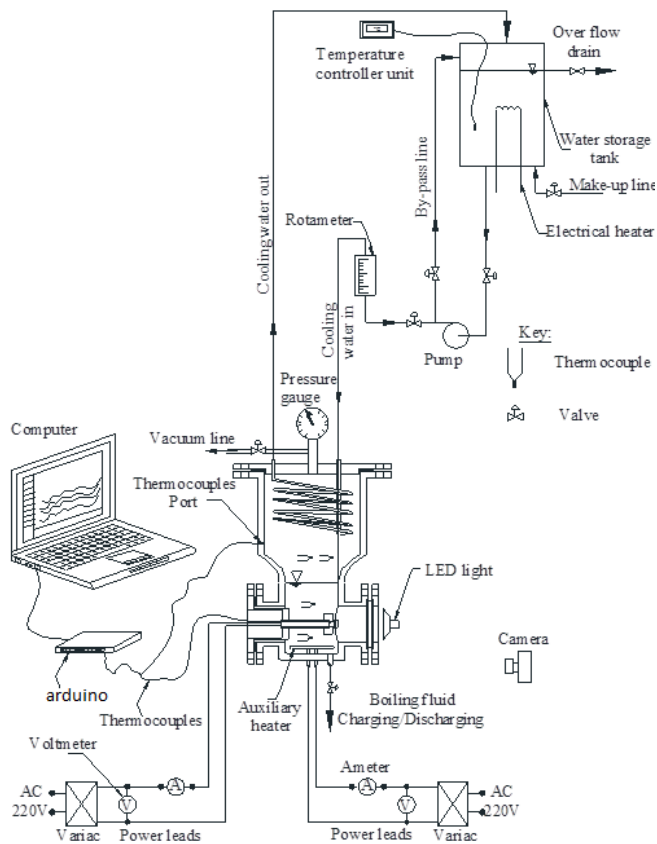

Fig. 3. Line diagram of the experimental test facility

The test tube assembly, shown in Fig. 4, consists of a horizontal copper tube of 24 mm outer diameter, 22 mm inner diameter, with a length of 125 mm. The test section inside the supply the input power in steps to the test tube cartridge heater, while the second is used to supply the power to the auxiliary heater. An Arduino MEGA board and laptop are used to monitor the measured temperatures. Eight calibrated copper constantan (T-type) thermocouples are connected to the boarding card. A parallax microcontroller add-on tool for data acquisition (PLX-DAQ) receives the output signal from all thermocouples and feeds them into a Microsoft Excel spreadsheet.

*D. Experimental preparations*

The test fluids are prepared by dissolving weighted additives samples in distilled water to produce the desired concentration (ppm). For the additive in powder form ($SiO_2$ and $Al_2O_3$), sample weights are measured using a precision electronic weighing machine of ± 0.1 mg accuracy. $SiO_2$ nanoparticles of 0.750, 1.5, 2.25, 3, and 3.75 g are weighted and added to a beaker containing 1 liter of distilled water. The solution is stirred with a magnetic stirring device. Thus the desired concentrations of 0.001, 0.002, 0.003, 0.005, and 0.01 vol% are achieved. The same procedures are followed to prepare the second nanoparticle ($Al_2O_3$) solution but with the following different weights 0.125, 0.250, 0.375, 0.500, and 0.625 g. Thus, the desired concentrations are achieved.

Eight calibrated copper – constantan thermocouples are incorporated within the test loop to monitor the temperature in various positions. Four are used to measure the surface temperature of the test tube. The test tube wall temperature is taken as the averaged value. The liquid pool temperature is monitored by two thermocouples placed 20 mm below the free surface of the liquid pool and 10 mm below the test tube. Another two thermocouples are used to monitor the vapor temperature. They are positioned midway along the test tube about 20 mm above the liquid pool surface. The outputs of all thermocouples are received and scanned by the Arduino control system interfaced with a laptop. A calibrated pressure gauge (-1: 5 bar), connected to the boiling and condensation vessel, is used to measure the operating pressure during the pool boiling experiments. Also, it is used to monitor the pressure of boiling and condensation vessels when tested to check the leakage.

Two thermostats with two certificated mercury thermometers are used to calibrate the copper-constantan thermocouples simultaneously using their hot water baths, as shown in Fig. 6. In this manner, the fluctuation in water temperature during the calibration process can be minimized. The calibration steps are started from the ice point to the boiling point of distilled water. It should be mentioned that the calibration of the thermocouple above boiling point has been carried out using oil heated stepwise gradually up to 140 °C.

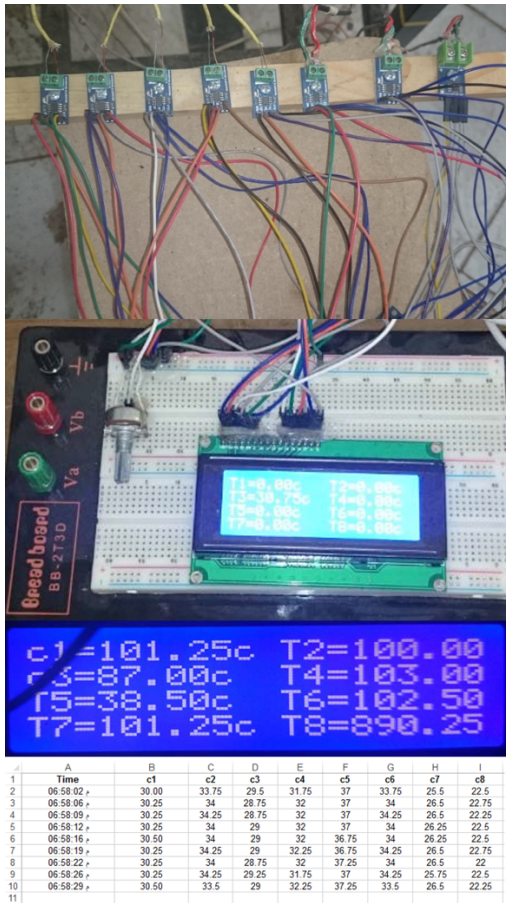
Fig. 5. Arduino MEGA board with eight thermocouples for temperature measurements.

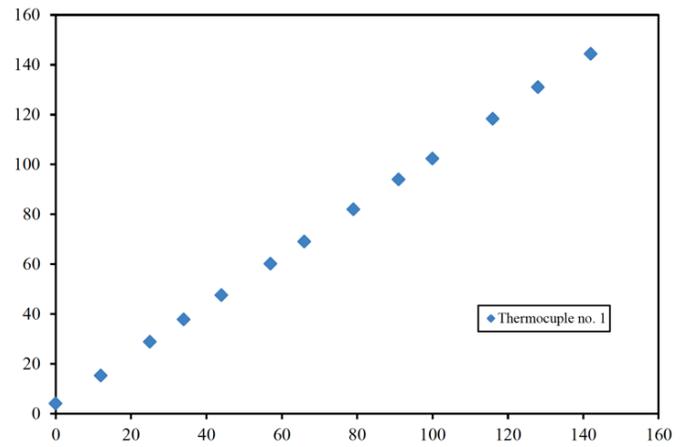
Fig. 6. Calibration curve of thermocouple no. 1.

The bourdon tube pressure gauge is calibrated using the deadweight tester within the investigated pressure range, as shown in Fig. 7. It is done by using known weights for changing the applied pressure on the gauge to be calibrated.

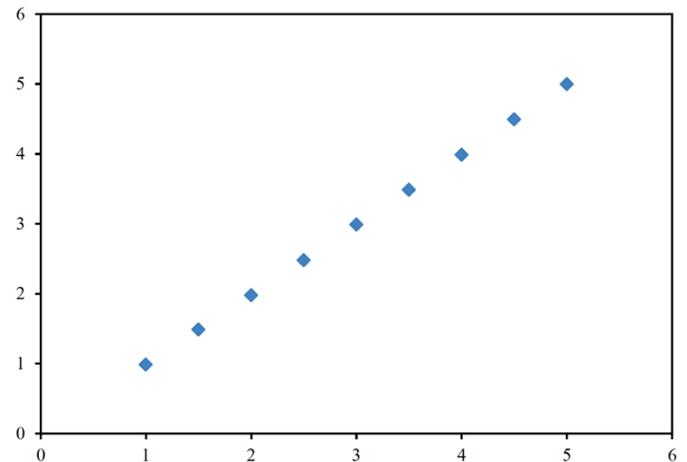
Fig. 7. Calibration curve of the pressure gauge.

## IV. Results

### A. Photographic visualization

The primary boiling control is directly related to nucleation site activation and density (influenced by wetting) and bubble dynamics (growth through departure, influenced by interfacial tension) that can be surmised from the photographic presentation in Fig. 8.

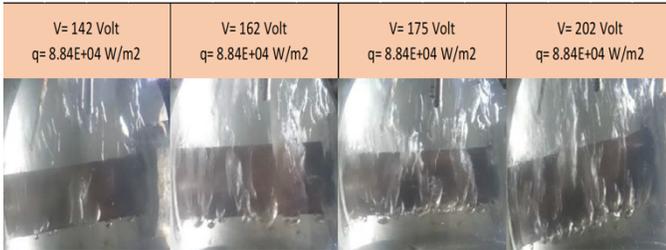

Fig. 8. Developing boiling steps for distilled water and aqueous SiO2 solutions of different concentrations, using smooth tube surface

### B. Repeatability of measurements

Fig. 9 shows the nucleate pool boiling data for pure distilled water using the smooth surface, and excellent repeatability is observed with runs 1 and 2, indicating stabilization of the nucleation cavities; these runs have been carried out. These data provide an accurate baseline reference for the nucleate boiling performance of the surfactant solutions and nanofluids.

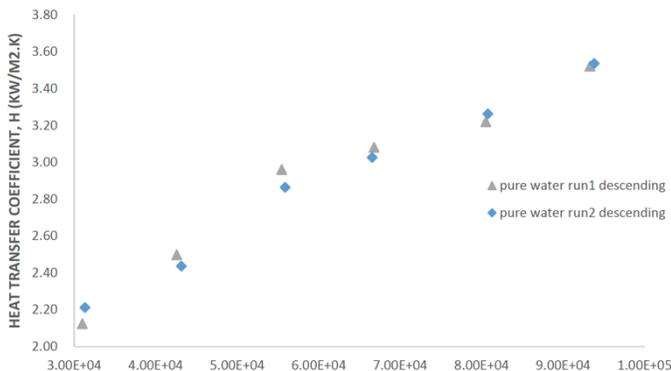

Fig. 9. CHF and HTC measurements for two experimental runs to checks its repeatability

### C. Surfactants

Figure 10 shows the comparative results of the variation of h with $q_w$ using pure water and different aqueous surfactants solutions (SDS and SLES).

This behavior is attributed to the agitation effect resulting from the mobility of the vapor bubbles emitted from the tube wall nucleation sites, which then travel through the liquid pool. The increase in $q_w$ activates a more significant number of nucleation sites. It should also be noticed that for any level of $q_w$, increasing the concentration of aqueous surfactant solution causes an appreciable increase in h. The surfactant SDS has shown consistent enhancements compared to SLES and distilled water over the whole heat flux span. Also, it is worth noting that the distilled water had higher heat transfer coefficient values till heat flux of 70kw/m**2**. Afterward, they tend to have the same values. An explanation for the observed enhancement in h could be given by considering the role of dynamic surface tension and the subsequent modification of bubble dynamics. With the nucleation of a vapor bubble and its subsequent growth, diffusion of surfactant molecules and their adsorption/desorption rates at the interface govern the extent of dynamic surface tension. The dynamic surface tension is appreciably lower than the solvent's surface tension, which helps to promote a large number of active nucleation sites. Lower values of the dynamic surface tension also allow the departure of smaller-sized bubbles because of the reduction in surface tension force at the heated tube wall that counters the bouncy force trying to pull the bubble away from the tube wall. The bubble growth time will consequently be expected and lead to an increase in bubble departure frequency.

### D. $SiO_2$

We have investigated the effect of heat flux with different concentrations for SiO2 on the HTC enhancements. Figure 11 shows that n inverse relationship of SiO2 concentrations with HTC. SiO2 concentration of 0.001% has a consistent high HTC value for all different heat flux values. These enhancements decrease with the increase of concentration, reaching its lowest value of 0.03%. However, they have a thresh hold for high concentrations, 0.01% and 0.03%, where enhancements are reversed. 0.03% had slightly high enhancements until heat flux 50 kW/m2. Afterward, the 0.01% concentration had a slight advantage over 0.03%. To drill through the concentrations, we have analyzed lower concentrations than 0.001% to investigate if low concentrations will result in better results.

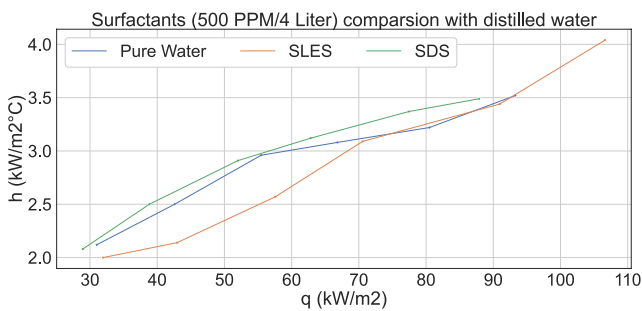

Fig. 10. Variation of h with $q_w$ for SLES and SDS at 500 PPM / 4Liter distilled water concentrations

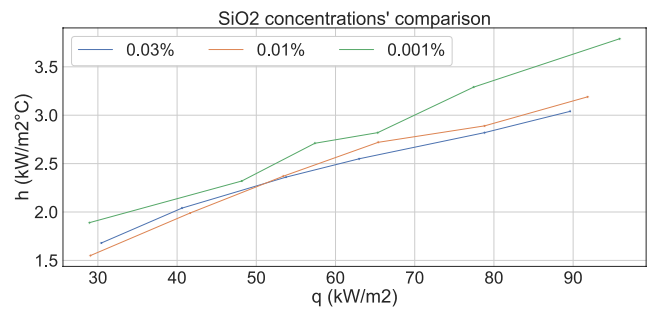

Fig. 11. Variation of h with $q_w$ for different SiO2 concentrations in distilled water

Concentrations below 0.001% showed extremely marginal differences. That confirms that Sio2 nanoparticles at concentration 0.001% has the highest HTC enhancements.

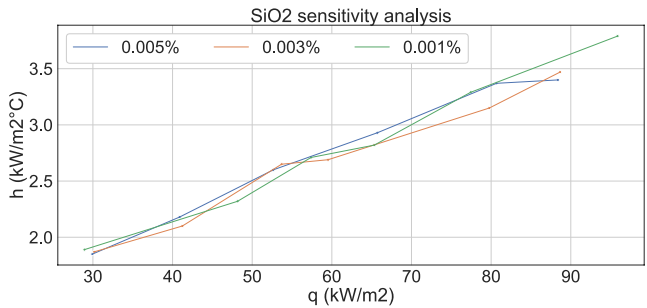

Fig. 12. Variation of h with $q_w$ for low SiO2 concentrations in distilled water

## E. Al2O3

Figure 13 shows a more apparent disparity between different concentrations compared to SiO2. However, the concentration with the highest HTC values is 0.01% compared to 0.001% with SiO2. Furthermore, the 0.03% has shown a higher slope compared to 0.01% that might surpass its HTC values with high heat flux. However, higher heat flux is practically unmanageable for our test rig due to the burn-out effect for our cylindrical heater. The disparity between the HTC values for 0.01% and 0.03% indicates other differences for concentrations in between. Therefore, further analysis for concentrations between 0.01% and 0.03% is needed to confirm if we can obtain different results.

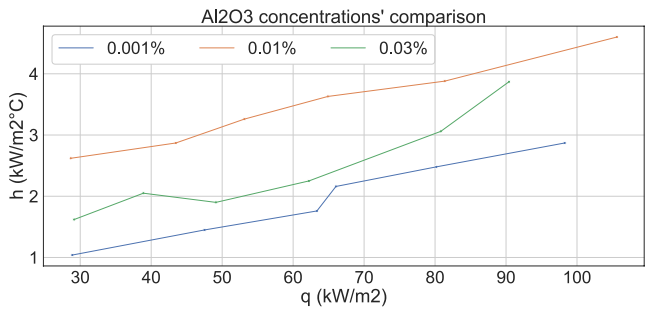

Fig. 13. Variation of h with $q_w$ for different Al2O3 concentrations in distilled water

For concentrations between 0.01% and 0.03, the 0.01% remains the highest HTC values over the whole heat flux data points. Additionally, the high trend for 0.02% and 0.03% remains remarkable at high heat flux values.

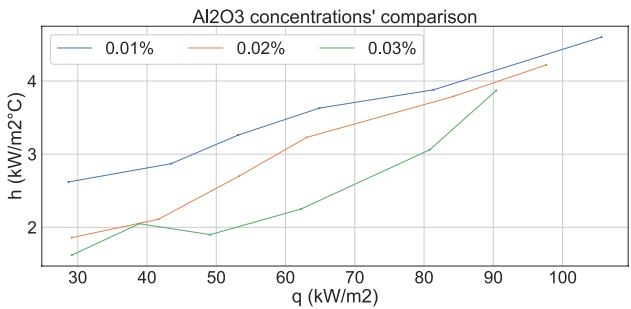

Fig. 14. Variation of h with $q_w$ for different Al2O3 concentrations in distilled water

## F. Comparison between SiO2, Al2O3, and distilled water

Figure 15 shows different graphs for SiO2, Al2O3, and distilled water at different concentrations. It indicates that at low concentrations, mainly 0.001%, SiO2 nanoparticles provide slightly better HTC values than distilled water and high enhancements than Al2O3. At 0.01%, the enhancements for Al2O3 are remarkably higher with a high slope. Such improvements are more evident at high heat flux compared to low values.

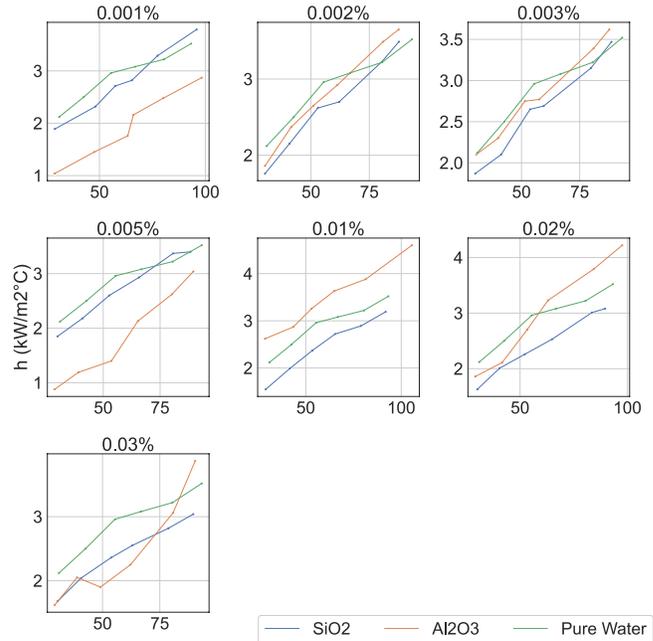

Fig. 15. Variation of h with $q_w$ for SiO2, Al2O3, and distilled water concentrations.

## V. CONCLUSION

From the above discussion of results, the following conclusions can be reached. Increasing the wall heat flux increases the nucleate pool boiling heat transfer coefficient (HTC) and the active nucleation site density for a given aqueous solution concentration and nanoparticles. Moreover, the enhancement in nucleate pool boiling depends on the wall heat flux, surfactant concentration, and nanoparticles. Secondly, the experimental results demonstrate that the heat transfer of the pool boiling process can be enhanced slightly by the addition of a small amount of surfactant in aqueous solutions and considerably by adding a definitive number of nanoparticles at high heat flux values. For a given wall heat flux, increasing the concentration up to a critical missile concentration ($C \leq c.m.c$) of aqueous SLES and SDS solutions increases both the nucleate pool boiling HTC and the active nucleation site density, while with ($C > c.m.c$), the enhancement decreases. For nanoparticles, SiO2 showed the highest enhancements at concentration 0.001% and Al2O3 at 0.01%. Al2O3 has shown a high slope at high heat flux values that might give higher HTC enhancements at high heat flux values.

## VI. REFERENCES

[1] Ciloglu D, Bolukbasi A. A comprehensive review on pool boiling of nano- fluids. Appl Therm Eng 2015;84:45–63.

[2] Kamatchi R, Venkatachalapathy S. Parametric study of pool boiling heat transfer with nanofluids for the enhancement of critical heat flux: a review. Int J Therm Sci 2015;87:228–40.

[3] Kshirsagar JM, Shrivastava R. Review of the influence of nanoparticles on thermal conductivity, nucleate pool boiling and critical heat flux. Heat Mass Transf 2015;51:381–98.

[4] Wu JM, Zhao J. A review of nanofluid heat transfer and critical heat flux enhancement – Research gap to engineering application. Prog Nucl Energy 2013;66:13–24.

[5] Celen A, Çebi A, Aktas M, Mahian O, Dalkilic AS, Wongwises S. A review of nanorefrigerants: flow characteristics and applications. Int J Refrig 2014;44:125–40.

[6] Bahiraei M, Hangi M. Flow and heat transfer characteristics of magnetic nanofluids: a review. J Magn Magn Mater 2015;374:125–38.

[7] Chopkar M, Das AK, Manna I, Das PK. Pool boiling heat transfer character- istics of ZrO2–water nanofluids from a flat surface in a pool. J Heat Mass Transf 2008;44:999–1004.

[8] Soltani S, Etemad SG, Thibault J. Pool boiling heat transfer performance of Newtonian nanofluids. Heat Mass Transf 2009;45:1555–60.

[9] Golubovic MN, Hettiarachchi HDM, Worek WM, Minkowycz WJ. Nanofluids and critical heat flux, experimental and analytical study. Appl. Therm. Eng 2009;29:1281–8.

[10] Huang CK, Lee CW, Wang CK. Boiling enhancement by TiO2 nanoparticle deposition. Int J Heat Mass Transf 2011;54:4895–903.

[11] Heris SZ. Experimental investigation of pool boiling characteristics of low- concentrated CuO/ethylene glycolewater nanofluids. Int Commun Heat Mass Transf 2011;38:1470–3.

[12] Sheikhbahai M, Nasr Esfahany M, Etesami N. Experimental investigation of pool boiling of Fe3O4/ethylene glycol-water nanofluid in electric field. Int J Therm Sci 2012;62:149–53.

[13] Pham QT, Kim TI, Lee SS, Chang SH. Enhancement of critical heat flux using nano-fluids for invessel retention-external vessel cooling. Appl Therm Eng 2012;35:157–65.

[14] Pham QT, Kim TI, Lee SS, Chang SH. Enhancement of critical heat flux using nano-fluids for invessel retention-external vessel cooling. Appl Therm Eng 2012;35:157–65.

[15] Sheikhbahai M, Esfahany MN, Etesami N. Experimental investigation of pool boiling of Fe3O4/ethylene glycol-water nanofluid in electric field. Int J Therm Sci 2012;62:149–53.

[16] Raveshi MR, Keshavarz A, Mojarrad MS, Amiri S. Experimental investigation of pool boiling heat transfer enhancement of alumina–water–ethylene glycol nanofluids. Exp Therm Fluid Sci 2013;44:805–14.

[17] Vazquez DM, Kumar R. Surface effects of ribbon heaters on critical heat flux in nanofluid pool boiling. Int Commun Heat Mass Transf 2013;41:1–9.

[18] Lee JH, Lee T, Jeong YH. The effect of pressure on the critical heat flux in water-based nanofluids containing Al2O3 and Fe3O4 nanoparticles. Int J Heat Mass Transf 2013;61:432–8.

[19] Shoghl SN, Bahrami M, Moraveji MK. Experimental investigation and CFD modeling of the dynamics of bubbles in nanofluid pool boiling. Int Commun Heat Mass Transf 2014;58:12–24.

[20] Tang X, Zhao YH, Diao YH. Experimental investigation of the nucleate pool boiling heat transfer characteristics of δ-Al2O3-R141b nanofluids on a hor- izontal plate. Exp Therm Fluid Sci 2014;52:88–96.

[21] Naphon P, Thongjing C. Pool boiling heat transfer characteristics of refrigerant-nanoparticle mixtures. Int Commun Heat Mass Transf 2014;52:84–9.

[22] Park SD, Moon SB, Bang IC. Effects of thickness of boiling-induced nano- particle deposition on the saturation of critical heat flux enhancement. Int J Heat Mass Transf 2014;78:506–14.

[23] Ulcay MS. CHF enhancement of Al2O3, TiO2 and Ag nanofluids and effect of nucleate pool boiling time. In: Proceedings of the 2014 IEEE intersociety conference on thermal and thermomechanical phenomena in electronic systems (ITherm); May 27–30, 2014. p. 756–64.

[24] Diao YH, Li CZ, Zhao YH, Liu Y, Wang S. Experimental investigation on the pool boiling characteristics and critical heat flux of Cu-R141b nanorefrigerant under atmospheric pressure Transf 2015;89:110–5.

[25] Umesh V, Raja B. A study on nucleate boiling heat transfer characteristics of pentane and CuO-pentane nanofluid on smooth and milled surfaces. Exp Therm Fluid Sci 2015;64:23–9.

[26] Kamatchi R, Venkatachalapathy S, Nithya C. Experimental investigation and mechanism of critical heat flux enhancement in pool boiling heat transfer with nanofluids. Heat Mass Transf 2015. http://dx.doi.org/10.1007/ s00231-015-1749-2.

[27] Sayahi T, Bahrami M. Investigation on the effect of type and size of nano- particles and surfactant on pool boiling heat transfer of nanofluids. J Heat Transf 2016;138(3):031502.

[28] Sayahi T, Bahrami M. Investigation on the effect of type and size of nano- particles and surfactant on pool boiling heat transfer of nanofluids. J Heat Transf 2016;138(3):031502.

[29] Sarafraz MM, Hormozi F. Comparatively experimental study on the boiling thermal performance of metal oxide and multi-walled carbon nanotube nanofluids. Powder Technol 2016;287:412–30.